\documentclass[12pt]{article}

\usepackage{url, graphicx, color, varioref, amsfonts, longtable, float}

\newtheorem{theorem}{Theorem}[section]
\newtheorem{lemma}[theorem]{Lemma}

\newenvironment{proof}[1][Proof]{\begin{trivlist}
\item[\hskip \labelsep {\bfseries #1}]}{\end{trivlist}}

\newcommand{\qed}{\nobreak \ifvmode \relax \else
      \ifdim\lastskip<1.5em \hskip-\lastskip
      \hskip1.5em plus0em minus0.5em \fi \nobreak
      \vrule height0.75em width0.5em depth0.25em\fi}

\title{A Hybrid Quantum Search Engine: A Fast Quantum Algorithm for Multiple Matches}

\author{Ahmed Younes\footnote {Birmingham, Edgbaston, B15 2TT, United Kingdom , axy@cs.bham.ac.uk} \,\,\,\,\,\,\, Jon Rowe \footnote {Birmingham, Edgbaston, B15 2TT, United Kingdom , jer@cs.bham.ac.uk} \\ School of Computer Science \\ University of Birmingham \\
\and
Julian Miller \footnote {York, Heslington, YO10 5DD, United Kingdom, jfm@ohm.york.ac.uk}\\  Department of Electronics\\   University of York\\  
}

\begin{document}
\maketitle
\begin{abstract}
In this paper we will present a quantum algorithm which works very efficiently in case of {\it multiple matches} 
within the search space and in the case of few matches, the algorithm performs 
classically. This allows us to propose a {\it hybrid quantum search engine} that integrates 
Grover's algorithm and the proposed algorithm here to have {\it general performance} 
better that any pure classical or quantum search algorithm.

\end{abstract}


\section{Introduction}

Quantum computers \cite{Deutsch85,Feynman86,Lloyd93} are probabilistic devices, which promise to do 
some types of computation more powerfully than classical computers \cite{Bern93,simon94}. 
Many quantum algorithms have been presented recently, for example, Shor \cite{shor97} presented a quantum 
algorithm for factorising a composite integer into its prime factors in polynomial time. 
Grover \cite{grover96} presented an algorithm for searching unstructured list of $N$ items with quadratic 
speed-up over algorithms run on classical computers.

Grover's algorithm inspired many researchers, including this work, to try to analyze and/or generalize 
his algorithm \cite{boyer96,Accardi00,Galindo00,Jozsa99,Bras00}. Grover's algorithm is proved to be optimal for a 
single match within the search space, although the number of iterations required by the algorithm increases; 
i.e. the problem becomes harder, as the number of matches exceeds half the number of items in the search space \cite{niel00}  
which is undesired behaviour for a search algorithm since the problem is expected to be easier. 
In this paper we will present a fast quantum algorithm, which can find a match among multiple matches 
within the search space after few iterations faster than any classical or quantum algorithm although 
for small number of matches the algorithm behaves classically. 


This leads us to proposing a hybrid search engine that includes Grover's 
algorithm and the algorithm proposed here. We also discuss the conditions that allow both algorithms to be integrate into a single hybrid.

The plan of the paper is as follows:  Section 2 gives a short introduction to quantum computers. 
Section 3 introduces the search problem and Grover's 
algorithm performance. Sections 4 to 6 introduce the proposed algorithm with analysis on its 
performance and behaviour. And we will end up with a conclusion in section 7.

\section{Quantum Computers}

\subsection{Quantum Bits}

In classical computers, a bit is considered as the basic unit for 
information processing; a bit can carry one value at a time (either 0 or 1). 
In quantum computers, the analogue of the bit is the quantum bit ({\it qubit} 
\cite{sch95}), which has two possible states encoded as $\left| 0 \right\rangle $ 
and $\left| 1 \right\rangle $; where the notation $\left| \,\,\, \right\rangle $ is 
called {\it Dirac Notation} and is considered as the standard notation of states in quantum 
mechanics \cite{dirac47}. For quantum computing purposes, the states $\left| 0 
\right\rangle $ and $\left| 1 \right\rangle$ can be considered as the 
classical bit values 0 and 1 respectively. An important difference between a 
classical bit and a qubit is that the qubit can exist in a linear 
superposition of both states ($\left| 0 \right\rangle $ and $\left| 1 
\right\rangle )$ at the same time and this gives the hope that quantum computers 
can do computation simultaneously ({\it Quantum Parallelism}). If we consider a quantum register with 
$n$ qubits all in superposition, then any operation applied on this register will be applied on 
the $2^{n}$ states representing the superposition simultaneously.

\subsection{Quantum Measurements}

To read information from a quantum register (quantum system), we must apply 
a measurement on that register which will result in a projection of the 
states of the system to a subspace of the state space compatible with the 
values being measured. For example, 
consider a two-qubit system $\left| \phi \right\rangle $ defined as follows:

\begin{equation}
\label{SATeq2}
\left| \phi \right\rangle = \alpha \left| {00} \right\rangle + \beta \left| 
{01} \right\rangle + \gamma \left| {10} \right\rangle + \delta \left| {11} 
\right\rangle ,
\end{equation}

\noindent
where $\alpha $, $\beta $, $\gamma $, and $\delta $ are complex numbers called the 
amplitudes of the system and satisfy $\left| \alpha \right|^2 + \left| \beta \right|^2 + \left| \gamma 
\right|^2 + \left| \delta \right|^2 = 1$. The probability that the first qubit of 
$\left| \phi \right\rangle $ to be $\left| 0 \right\rangle $ is equal to 
$\left( {\left| \alpha \right|^2 + \left| \beta \right|^2} \right)$. 
If for some reasons we need to have the value $\left| 0 \right\rangle $ 
in the first qubit after any measurement, we must try some how 
to increase its probability before applying the measurement. Note that, the new 
state after applying measurement must be re-normalized so the total probability is 
still 1.

\subsection{Quantum Gates}

In general, quantum algorithms can be understood as follows: Apply a series 
of transformations (gates) then apply the measurement to get the desired 
result with high probability. According to the laws of quantum mechanics and to keep the 
reversibility condition required in quantum computation, the evolution of 
the state of the quantum system $\left| \psi \right\rangle $ of size $n$ by 
time $t$ is described by a matrix $U$ of dimension $2^n\times 
2^n$ \cite{niel00}:

\begin{equation}
\label{SATeq4}
\left| {\psi '} \right\rangle = U\left| \psi \right\rangle ,
\end{equation}

\noindent
where $U$ satisfies the unitary condition: $U^{\dag} U = I$, where $U^{\dag}$ denotes 
the complex conjugate transpose of $U$ and $I$ is the identity matrix. For example, the $X$ gate ($NOT$ gate) is 
a single qubit gate (single input/output) similar in its effect to the classical $NOT$ gate. It inverts the 
state $\left| 0 \right\rangle $ to the state $\left| 1 \right\rangle$ and 
visa versa. It's $2\times2$ unitary matrix takes this form,

\begin{equation}
\label{SATeq5}
X = \left[ {{\begin{array}{*{20}c}
 0 \hfill & 1 \hfill \\
 1 \hfill & 0 \hfill \\
\end{array} }} \right],
\end{equation}

\noindent
and its circuit takes the form shown in Fig.(\ref{SATfig1}). Notice that, from now on we assume that 
a horizontal line used in a quantum circuit represents a qubit and the flow of the circuit logic 
is from left to right. For circuits with multiple qubits, qubits will be arranged according to 
the notation used in the figure.

\begin{center}
\begin{figure}[H]
\begin{center}
\setlength{\unitlength}{3947sp}%
\begingroup\makeatletter\ifx\SetFigFont\undefined%
\gdef\SetFigFont#1#2#3#4#5{%
  \reset@font\fontsize{#1}{#2pt}%
  \fontfamily{#3}\fontseries{#4}\fontshape{#5}%
  \selectfont}%
\fi\endgroup%
\begin{picture}(1875,399)(4201,-2998)
\thinlines
{\color[rgb]{0,0,0}\put(4701,-2986){\framebox(600,375){}}
}%
{\color[rgb]{0,0,0}\put(5301,-2761){\line( 1, 0){300}}
}%
{\color[rgb]{0,0,0}\put(4701,-2761){\line(-1, 0){300}}
}%
\put(4926,-2836){$X$}%
\put(3101,-2836){$\left( {\alpha \left| 0 \right\rangle  + \beta \left| 1 \right\rangle } \right)$}%
\put(5676,-2836){$\left( {\beta \left| 0 \right\rangle  + \alpha \left| 1 \right\rangle } \right)$}%
\end{picture}
\end{center}
\caption{$NOT$ gate quantum circuit.}
\label{SATfig1}
\end{figure}
\end{center}

Another important example is the Hadamard gate ($H$ gate) which has no classical equivalent; it produces 
a completely random output with equal probabilities of the output to be 
$\left| 0 \right\rangle $ or $\left| 1 \right\rangle $ on any measurements. 
It's $2\times2$ unitary matrix takes this form,

\begin{equation}
\label{SATeq6}
H = \frac{1}{\sqrt 2 }\left[ {{
\begin{array}{*{20}c}
 1 \hfill & \,\,\,\,1 \hfill \\
 1 \hfill & { - 1} \hfill \\
\end{array}
}} \right],
\end{equation}

\noindent
and its circuit takes the form shown in Fig.(\ref{SATfig2}).

\begin{center}
\begin{figure}[H]
\begin{center}
\setlength{\unitlength}{3947sp}%
\begingroup\makeatletter\ifx\SetFigFont\undefined%
\gdef\SetFigFont#1#2#3#4#5{%
  \reset@font\fontsize{#1}{#2pt}%
  \fontfamily{#3}\fontseries{#4}\fontshape{#5}%
  \selectfont}%
\fi\endgroup%
\begin{picture}(1875,399)(4201,-2998)
\thinlines
{\color[rgb]{0,0,0}\put(4301,-2986){\framebox(600,375){}}
}%
{\color[rgb]{0,0,0}\put(4901,-2761){\line( 1, 0){300}}
}%
{\color[rgb]{0,0,0}\put(4301,-2761){\line(-1, 0){300}}
}%
\put(4526,-2836){$H$}%
\put(3750,-2836){$\left| x \right\rangle$}%
\put(5276,-2836){$\frac{1}{{\sqrt 2 }}\left( {\left| 0 \right\rangle  + \left( { - 1} \right)^x \left| 1 \right\rangle } \right)$}%
\end{picture}
\end{center}
\caption{Hadamard gate quantum circuit, where $x$ is any Boolean variable.}
\label{SATfig2}
\end{figure}
\end{center}

Controlled operations are considered as the heart of quantum computing \cite{bare95}, the Controlled-$U$ gate is the 
general case for any controlled gate with one or more control qubit(s) as shown in Fig.(\ref{SATfig3_4_X}.a). It works as 
follows: If any of the control qubits $\left| {c_i } \right\rangle $'s ($1 \le i \le n - 1)$ is 
set to 0, then the quantum gate $U$ will not be applied on target qubit $\left| t \right\rangle $; i.e. 
$U$ is applied on $\left| t \right\rangle$ if and only if all $\left| {c_i } \right\rangle $'s are set to 1. 
The states of the qubits after applying the gate will be transformed according to the following rule:

\begin{equation}
\label{SATeq7}
\begin{array}{l}
 \left| {c_i } \right\rangle \to \left| {c_i } \right\rangle ;1 \le i \le n - 1\\ 
 \left| t \right\rangle \to \left| t_{CU}\right\rangle = U^{c_1 c_2 ...c_{n - 1} }\left| t \right\rangle 
\\ 
 \end{array}
\end{equation}

\noindent
where $c_1 c_2 ...c_{n - 1}$ in the exponent of $U$ means the $AND$-ing of the 
qubits $c_1 ,\,c_2 ,...,c_{n - 1} $.


\begin{center}
\begin{figure} [H]
\begin{center}
\setlength{\unitlength}{3947sp}%
\begingroup\makeatletter\ifx\SetFigFont\undefined%
\gdef\SetFigFont#1#2#3#4#5{%
  \reset@font\fontsize{#1}{#2pt}%
  \fontfamily{#3}\fontseries{#4}\fontshape{#5}%
  \selectfont}%
\fi\endgroup%
\begin{picture}(3525,1485)(3376,-1936)
{\color[rgb]{0,0,0}\thinlines
\put(4201,-1261){\circle*{150}}
}%
{\color[rgb]{0,0,0}\put(4201,-586){\circle*{150}}
}%
{\color[rgb]{0,0,0}\put(4201,-811){\circle*{150}}
}%
{\color[rgb]{0,0,0}\put(6301,-586){\circle*{150}}
}%
{\color[rgb]{0,0,0}\put(6301,-811){\circle*{150}}
}%
{\color[rgb]{0,0,0}\put(6301,-1261){\circle*{150}}
}%
{\color[rgb]{0,0,0}\put(6301,-1561){\circle{150}}
}%
{\color[rgb]{0,0,0}\put(4201,-811){\circle*{150}}
}%
{\color[rgb]{0,0,0}\put(4201,-586){\circle*{150}}
}%
{\color[rgb]{0,0,0}\put(4201,-586){\circle{150}}
}%
{\color[rgb]{0,0,0}\put(4051,-1711){\framebox(300,300){}}
}%
{\color[rgb]{0,0,0}\put(4351,-1561){\line( 1, 0){300}}
}%
{\color[rgb]{0,0,0}\put(4051,-1561){\line(-1, 0){300}}
}%
{\color[rgb]{0,0,0}\put(4201,-1111){\line( 0,-1){300}}
}%
{\color[rgb]{0,0,0}\put(4651,-586){\line(-1, 0){900}}
}%
{\color[rgb]{0,0,0}\put(4201,-511){\line( 0,-1){450}}
}%
{\color[rgb]{0,0,0}\put(4651,-811){\line(-1, 0){900}}
}%
{\color[rgb]{0,0,0}\put(4651,-1261){\line(-1, 0){900}}
}%
{\color[rgb]{0,0,0}\put(6751,-586){\line(-1, 0){900}}
}%
{\color[rgb]{0,0,0}\put(6751,-811){\line(-1, 0){900}}
}%
{\color[rgb]{0,0,0}\put(6751,-1261){\line(-1, 0){900}}
}%
{\color[rgb]{0,0,0}\put(6751,-1561){\line(-1, 0){900}}
}%
{\color[rgb]{0,0,0}\put(6301,-511){\line( 0,-1){450}}
}%
{\color[rgb]{0,0,0}\put(6301,-1111){\line( 0,-1){525}}
}%

\put(4185,-1120){$\vdots$}
\put(6285,-1120){$\vdots$}

\put(4801,-586){$\left| {c_1 } \right\rangle$}%
\put(3276,-586){$\left| {c_1 } \right\rangle$}%
\put(5376,-586){$\left| {c_1 } \right\rangle$}%
\put(6901,-586){$\left| {c_1 } \right\rangle$}%

\put(4801,-886){$\left| {c_2 } \right\rangle$}%
\put(3276,-886){$\left| {c_2 } \right\rangle$}%
\put(5376,-886){$\left| {c_2 } \right\rangle$}%
\put(6901,-886){$\left| {c_2 } \right\rangle$}%

\put(4801,-1561){$\left| {t_{CU} } \right\rangle$}%
\put(3276,-1561){$\left| {t } \right\rangle$}%
\put(5376,-1561){$\left| {t } \right\rangle$}%
\put(6901,-1561){$\left| {t_{CN} } \right\rangle$}%

\put(4801,-1261){$\left| {c_{n-1} } \right\rangle$}%
\put(3276,-1261){$\left| {c_{n-1} } \right\rangle$}%
\put(5376,-1261){$\left| {c_{n-1} } \right\rangle$}%
\put(6901,-1261){$\left| {c_{n-1} } \right\rangle$}%

\put(4126,-1636){$U$}%

\put(5776,-1936){b.Controlled-NOT}%
\put(3700,-1936){a.Controlled-U}%
\end{picture}
\end{center}
\caption{Controlled gates where the back circle $\bullet $ indicates 
the control qubits, and the symbol $ \oplus $ in part (b.) indicates the target qubit.}
\label{SATfig3_4_X}
\end{figure}
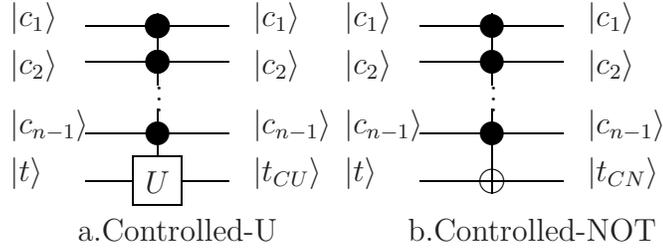
\end{center}

If $U$ in the general Controlled-$U$ gate is replaced with the $X$ gate mentioned above, the resulting gate is  
called a Controlled-$NOT$ gate (shown in Fig.(\ref{SATfig3_4_X}.b)). It works as follows: It 
inverts the target qubit if and only if all the control qubits are set to 1. 
Thus the qubits of the system $c_1 ,c_2 ,...,c_{n - 1} ,t $ will be transformed 
according to the following rule:

\begin{equation}
\label{SATeq8}
\begin{array}{l}
 \left| {c_i } \right\rangle \to \left| {c_i } \right\rangle ;1 \le i \le n - 1 \\ 
 \left| {t } \right\rangle \to \left| t_{CN}\right\rangle = \left| {t \oplus c_1 c_2 ...c_{n - 1} } 
\right\rangle \\ 
 \end{array}
\end{equation}

\noindent
where $c_1 c_2 \ldots c_{n - 1} $ is the $AND$-ing of the qubits $c_1 ,c_2 
,\ldots ,c_{n - 1} $ and $ \oplus $ is the classical XOR operation.

\section{Search Problem}

Consider a list $L$ of $N$ items; $L = \{ 0,1,...,N - 1\}$, and consider a function $f$ 
which maps the items in $L$ to either 0 or 1 according to some properties these items shall satisfy; 
i.e. $f:L \to \{ 0,1\}$. The problem is to find any $i \in L$ such that $f(i) = 1$ assuming that 
such $i$ must exist in the list. It was shown classically that we need approximately 
${N \mathord{\left/ {\vphantom {N 2}} \right. \kern-\nulldelimiterspace} 2}$ tests to get a result with 
probability at least one-half. Let $M$ denotes the number of matches within the search space such that 
$1 \le M \le N$ and for simplicity and without loss of generality we can assume that $N = 2^n$. 
Grover's algorithm was shown to solve this problem \cite{boyer96} in $O\left( {\sqrt {N/M} } \right)$. 
In \cite{niel00}, it was shown that the number of iterations will increase 
for $M > N/2$ which is undesired behaviour for a search algorithm. To overcome this problem it was proposed 
in \cite{niel00} that the search space can be doubled so the number of matches is always less than half the search space and then iterate 
the algorithm $\pi /4\sqrt {2N/M}$ times so the algorithm still runs in $O\left( {\sqrt {N/M} } \right)$. 
But using this approach will double the cost of space/time requirement. In the following section we will present 
an algorithm that can find a solution for $M > N/2$ with probability at least $92.6\%$ after applying the algorithm 
once.

\section{The Algorithm}
\subsection{Iterating the algorithm once}

\begin{center}
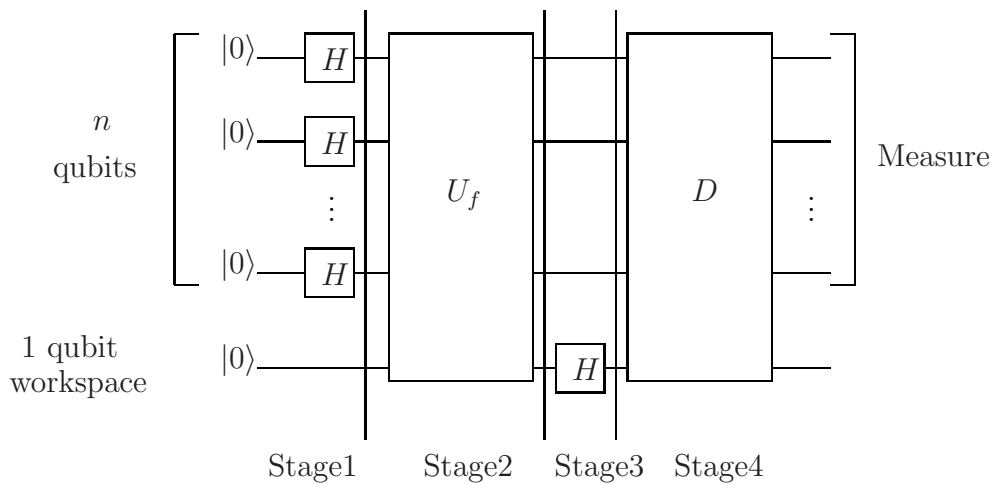
\begin{figure} 
\begin{center}
\setlength{\unitlength}{3947sp}%
\begingroup\makeatletter\ifx\SetFigFont\undefined%
\gdef\SetFigFont#1#2#3#4#5{%
  \reset@font\fontsize{#1}{#2pt}%
  \fontfamily{#3}\fontseries{#4}\fontshape{#5}%
  \selectfont}%
\fi\endgroup%
\begin{picture}(4950,2991)(2551,-3040)
\thinlines
{\color[rgb]{0,0,0}\put(3676,-361){\line( 1, 0){300}}
}%
{\color[rgb]{0,0,0}\put(3976,-511){\framebox(300,300){}}
}%
{\color[rgb]{0,0,0}\put(4276,-361){\line( 1, 0){225}}
}%
{\color[rgb]{0,0,0}\put(3676,-886){\line( 1, 0){300}}
}%
{\color[rgb]{0,0,0}\put(3976,-1036){\framebox(300,300){}}
}%
{\color[rgb]{0,0,0}\put(4276,-886){\line( 1, 0){225}}
}%
{\color[rgb]{0,0,0}\put(3976,-1861){\framebox(300,300){}}
}%
{\color[rgb]{0,0,0}\put(3676,-1711){\line( 1, 0){300}}
}%
{\color[rgb]{0,0,0}\put(4276,-1711){\line( 1, 0){225}}
}%
{\color[rgb]{0,0,0}\put(3676,-2311){\line( 1, 0){825}}
}%
{\color[rgb]{0,0,0}\put(4351,-61){\line( 0,-1){2700}}
}%
{\color[rgb]{0,0,0}\put(4501,-2386){\framebox(900,2175){}}
}%
{\color[rgb]{0,0,0}\put(5401,-361){\line( 1, 0){600}}
}%
{\color[rgb]{0,0,0}\put(5401,-886){\line( 1, 0){600}}
}%
{\color[rgb]{0,0,0}\put(5401,-1711){\line( 1, 0){600}}
}%
{\color[rgb]{0,0,0}\put(5401,-2311){\line( 1, 0){150}}
}%
{\color[rgb]{0,0,0}\put(5476,-61){\line( 0,-1){2700}}
}%
{\color[rgb]{0,0,0}\put(5551,-2461){\framebox(300,300){}}
}%
{\color[rgb]{0,0,0}\put(5851,-2311){\line( 1, 0){150}}
}%
{\color[rgb]{0,0,0}\put(5926,-61){\line( 0,-1){2700}}
}%
{\color[rgb]{0,0,0}\put(6001,-2386){\framebox(900,2175){}}
}%
{\color[rgb]{0,0,0}\put(6901,-361){\line( 1, 0){375}}
}%
{\color[rgb]{0,0,0}\put(6901,-886){\line( 1, 0){375}}
}%
{\color[rgb]{0,0,0}\put(6901,-1711){\line( 1, 0){375}}
}%
{\color[rgb]{0,0,0}\put(6901,-2311){\line( 1, 0){375}}
}%
{\color[rgb]{0,0,0}\put(3151,-211){\line( 0,-1){1575}}
}%
{\color[rgb]{0,0,0}\put(3151,-211){\line( 1, 0){150}}
}%
{\color[rgb]{0,0,0}\put(3151,-1786){\line( 1, 0){150}}
}%
{\color[rgb]{0,0,0}\put(7426,-211){\line( 0,-1){1575}}
}%
{\color[rgb]{0,0,0}\put(7276,-211){\line( 1, 0){150}}
}%
{\color[rgb]{0,0,0}\put(7276,-1786){\line( 1, 0){150}}
}%
{\color[rgb]{0,0,0}\put(7126,-1386){$\vdots$}}
{\color[rgb]{0,0,0}\put(4050,-1386){$\vdots$}}
\put(3376,-361){$\left| 0 \right\rangle$}%
\put(4010,-436){$H$}%
\put(4010,-961){$H$}%
\put(3376,-886){$\left| 0 \right\rangle$}%
\put(4010,-1786){$H$}%
\put(3376,-1711){$\left| 0 \right\rangle$}%
\put(3376,-2311){$\left| 0 \right\rangle$}%
\put(4795,-1261){$U_f$}%
\put(5585,-2386){$H$}%
\put(6335,-1261){$D$}%
\put(4651,-2986){Stage2}%
\put(3676,-2986){Stage1}%
\put(5476,-2986){Stage3}%
\put(6226,-2986){Stage4}%
\put(7501,-1036){Measure}%
\put(2576,-811){$n$}%
\put(2326,-1111){qubits}%
\put(2126,-2236){1 qubit}%
\put(2051,-2461){workspace}%
\end{picture}
\end{center}
\caption{Quantum circuit for the proposed algorithm.}
\label{SATfig6}
\end{figure}
\end{center}

For a list of size $N=2^n$, the steps of the algorithm can be understood 
as follows as shown in Fig.(\ref{SATfig6}): 

\begin{itemize}
\item[1-]{\it Register Preparation}. Prepare a quantum register of $n+1$ 
qubits all in state $\left| 0 \right\rangle $, where the extra qubit is 
used as a workspace for evaluating the oracle $U_f$:

\begin{equation}
\label{SATeq10}
\left| {W_0 } \right\rangle = \left| 0 \right\rangle ^{ \otimes n} \otimes 
\left| 0 \right\rangle. 
\end{equation}

\item[2-] {\it Register Initialization}. Apply Hadamard gate on each of the first $n$ qubits in parallel, 
so they contain the $2^{n}$ states, where $i$ is the integer representation of items in the list:

\begin{equation}
\label{SATeq11}
\left| {W_1 } \right\rangle = \left( {H^{ \otimes n} \otimes I} 
\right)\left| {W_0 } \right\rangle = \left( {\frac{1}{\sqrt N 
}\sum\limits_{i = 0}^{N - 1} {\left| i \right\rangle } } \right) \otimes 
\left| 0 \right\rangle ;\,\,N = 2^n.
\end{equation}

\item[3-] {\it Applying Oracle}. Apply the oracle $U_{f}$ to map the items in the list to either 0 or 1
simultaneously and store the result in the extra workspace qubit:

\begin{equation}
\label{SATeq12}
\left| {W_2 } \right\rangle = U_f \left| {W_1 } \right\rangle = 
\frac{1}{\sqrt N }\sum\limits_{i = 0}^{N - 1} {\left( {\left| i 
\right\rangle \otimes \left| {0 \oplus f(i)} \right\rangle } \right)} = 
\frac{1}{\sqrt N }\sum\limits_{i = 0}^{N - 1} {\left( {\left| i 
\right\rangle \otimes \left| {f(i)} \right\rangle } \right)}. 
\end{equation}


\item[4-] {\it Completing Superposition and Changing Sign}. Apply Hadamard gate on the 
workspace qubit. This will extend the superposition for the $n+1$ qubits with 
the amplitudes of the desired states with negative sign as follows:

\begin{equation}
\label{SATeq13}
\begin{array}{l}
\left| {W_3 } \right\rangle = \left( {I^{ \otimes n} \otimes H} 
\right)\left| {W_2 } \right\rangle = \frac{1}{\sqrt N }\sum\limits_{i = 
0}^{N - 1} {\left( {\left| i \right\rangle \otimes \left( {\frac{\left| 0 
\right\rangle + ( - 1)^{f(i)}\left| 1 \right\rangle }{\sqrt 2 }} \right)} 
\right)}\\ 
\,\,\,\,\,\,\,\,\,\,\,\,\, = \frac{1}{\sqrt P }\sum\limits_{i = 0}^{N - 1} {\left( {\left| i 
\right\rangle \otimes \left( {\left| 0 \right\rangle + ( - 1)^{f(i)}\left| 1 
\right\rangle } \right)} \right)} ;\,\,\,P = 2N = 2^{n + 1}.\\
\end{array}
\end{equation}

Let $M$ be the number of matches, which makes the 
oracle $U_f$ evaluate to 1 (solutions); such that $1 \le M \le N$; assume that 
$\sum\nolimits_i {^{'}} $ indicates a sum over all $i$ which are desired matches ($2M$ states), 
and $\sum\nolimits_i {^{''}} $ indicates a sum over all $i$ which 
are undesired items in the list. So, $\left| {W_3 } \right\rangle $ can be re-written as follows:

\begin{equation}
\label{SATeq14}
\begin{array}{l}
\left| {W_{3} } \right\rangle  = \frac{1}{{\sqrt P }}\sum\limits_{i = 0}^{N - 1} {^{'} \left( {\left| i \right\rangle  \otimes \left( {\left| 0 \right\rangle  - \left| 1 \right\rangle } \right)} \right)}  \\ 
\,\,\,\,\,\,\,\,\,\,\,\,\,\,\,\,\,\, + \frac{1}{{\sqrt P }}\sum\limits_{i = 0}^{N - 1} {^{''} \left( {\left| i \right\rangle  \otimes \left( {\left| 0 \right\rangle  + \left| 1 \right\rangle } \right)} \right)}  \\ 
\,\,\,\,\,\,\,\,\,\,\,\,\,\,\, = \frac{1}{\sqrt P }\sum\limits_{i = 0}^{N - 1} 
{^{'}\left( {\left| i \right\rangle \otimes \left| 0 \right\rangle } \right)} 
- \frac{1}{\sqrt P }\sum\limits_{i = 0}^{N - 1} {^{'}\left( {\left| i 
\right\rangle \otimes \left| 1 \right\rangle } \right)} \\ 
\,\,\,\,\,\,\,\,\,\,\,\,\,\,\,\,\,\, + \frac{1}{\sqrt P }\sum\limits_{i = 0}^{N - 
1} {^{''}\left( {\left| i \right\rangle \otimes \left| 0 \right\rangle } 
\right)} + \frac{1}{\sqrt P }\sum\limits_{i = 0}^{N - 1} {^{''}\left( 
{\left| i \right\rangle \otimes \left| 1 \right\rangle } \right)}. \\ 
 \end{array}
\end{equation}

From Eqn.(\ref{SATeq14}); we can see that there are $M$ states with amplitude 
$\left( {{\raise0.5ex\hbox{$\scriptstyle -1$}\kern-0.1em/\kern-0.15em\lower0.25ex\hbox{$\scriptstyle {\sqrt P }$}}} \right)$ 
where $f(i)=1$, and $(P-M)$ states with amplitude 
$\left( {{\raise0.5ex\hbox{$\scriptstyle 1$}\kern-0.1em/\kern-0.15em\lower0.25ex\hbox{$\scriptstyle {\sqrt P }$}}} \right)$. 
Notice that, applying Hadamard gate on the extra qubit splits the $\left| i \right\rangle$ 
states (solution states), to $M$ states 
($\sum\nolimits_{i}{^{'}\left( {\left| i \right\rangle \otimes \left| 0 \right\rangle } \right)}$) 
with positive amplitude 
$\left( {{\raise0.5ex\hbox{$\scriptstyle 1$}\kern-0.1em/\kern-0.15em\lower0.25ex\hbox{$\scriptstyle {\sqrt P }$}}} \right)$ 
and $M$ states 
($\sum\nolimits_{i}{^{'}\left( {\left| i \right\rangle \otimes \left| 1 \right\rangle } \right)}$) 
with negative amplitude $\left( {{\raise0.5ex\hbox{$\scriptstyle -1$}\kern-0.1em/\kern-0.15em\lower0.25ex\hbox{$\scriptstyle {\sqrt P }$}}} \right)$.

\item[5-]{\it Inversion About the Mean}. Apply the {\it Diffusion Operator} $D$ similar to 
that used in Grover's algorithm \cite{grover96} on the $n+1$ 
qubits. The diagonal representation of the diffusion operator $D$ can take this form:

\begin{equation}
\label{SATeqn16}
D = H^{ \otimes n + 1}\left( {2\left| 0 \right\rangle \left\langle 0 \right| 
- I} \right)H^{ \otimes n + 1} = 2\left| \psi \right\rangle \left\langle 
\psi \right| - I.
\end{equation}

\noindent
where, $\left| \psi \right\rangle = \frac{1}{\sqrt P }\sum\nolimits_{k = 0}^{P 
- 1} {\left| k \right\rangle }$  is an equally weighted 
superposition of states. The effect of applying $D$ \cite{niel00} on a general state 
$\sum\nolimits_{k = 0}^{P-1} {\alpha _k \left| k \right\rangle }$ produces 
$\sum\nolimits_{k = 0}^{P-1} {\left[ { - \alpha _k + 2\left\langle \alpha 
\right\rangle } \right]\left| k \right\rangle }$, where, $\left\langle \alpha \right\rangle = \frac{1}{P}\sum\nolimits_{k = 0}^{P-1} 
{\alpha _k }$ is the mean of the amplitudes of all states in the superposition; i.e. the amplitudes $\alpha _k $ will be transformed 
according to the following relation:

\begin{equation}
\label{SATeqn17}
\alpha _k \to \left[ { - \alpha _k + 2\left\langle \alpha \right\rangle } 
\right].
\end{equation}

In  our case, there are $M$ states with amplitude 
$\left( {{\raise0.5ex\hbox{$\scriptstyle -1$}\kern-0.1em/\kern-0.15em\lower0.25ex\hbox{$\scriptstyle {\sqrt P }$}}} \right)$ 
and $P-M$ states with amplitude
$\left( {{\raise0.5ex\hbox{$\scriptstyle 1$}\kern-0.1em/\kern-0.15em\lower0.25ex\hbox{$\scriptstyle {\sqrt P }$}}} \right)$, 
so the mean $\left\langle \alpha  \right\rangle$ is as follows:

\begin{equation}
\label{SATmean}
\left\langle \alpha \right\rangle = \frac{1}{P}\left( 
{M\left( {\frac{ - 1}{\sqrt P }} \right) + (P - M)\left( {\frac{1}{\sqrt P 
}} \right)} \right).
\end{equation}

So, applying $D$ on the system $\left| {W_{3}}\right\rangle$ shown in Eqn.(\ref{SATeq14}) 
can be understood as follows:

\begin{itemize}

\item[a-] The $M$ negative sign amplitudes (solutions): will be transformed from 
$\left( {{\raise0.5ex\hbox{$\scriptstyle -1$}
\kern-0.1em/\kern-0.15em
\lower0.25ex\hbox{$\scriptstyle {\sqrt P }$}}} \right)$ to $a$ , 
where $a$ is calculated as follows: Substitute $\alpha _k = \frac{ 
- 1}{\sqrt P }$ and $\left\langle \alpha  \right\rangle$ shown (Eqn.(\ref{SATmean})) in Eqn.(\ref{SATeqn17}) we get:

\begin{equation}
\label{SATeqn18}
\begin{array}{l}
 a = - \left( {\frac{ - 1}{\sqrt P }} \right) + \frac{2}{P}\left( {M\left( 
{\frac{ - 1}{\sqrt P }} \right) + (P - M)\left( {\frac{1}{\sqrt P }} 
\right)} \right) \\ 
 \,\,\,\,\, = \frac{1}{\sqrt P }\left( {3 - \frac{4M}{P}} \right). \\ 
 \end{array}
\end{equation}

\item[b-] The $(P-M)$ positive sign amplitudes will be transformed from 
$\left( {{\raise0.5ex\hbox{$\scriptstyle 1$}
\kern-0.1em/\kern-0.15em
\lower0.25ex\hbox{$\scriptstyle {\sqrt P }$}}} \right)$ 
to $b$ , where 
$b$ is calculated as follows: Substitute $\alpha _k = \frac{1}{\sqrt P }$ and $\left\langle \alpha  \right\rangle$ 
shown (Eqn.(\ref{SATmean})) in Eqn. (\ref{SATeqn17}) we get:

\begin{equation}
\label{SATeqn19}
\begin{array}{l}
 b = - \left( {\frac{1}{\sqrt P }} \right) + \frac{2}{P}\left( {M\left( 
{\frac{ - 1}{\sqrt P }} \right) + (P - M)\left( {\frac{1}{\sqrt P }} 
\right)} \right) \\ 
 \,\,\,\,\,\,\,\, = \frac{1}{\sqrt P }\left( {1 - \frac{4M}{P}} \right). \\ 
 \end{array}
\end{equation}
\end{itemize}

We can see that $a>b$ after applying $D$. 
The new system $\left| {W_{4}} \right\rangle$ can be written as follows:

\begin{equation}
\label{SATeqn20}
\begin{array}{l}
 D\left| {W_{3} } \right\rangle = \left| {W_{4} } \right\rangle = 
b \sum\limits_{i = 0}^{N - 1} {^{'}\left( {\left| i \right\rangle \otimes 
\left| 0 \right\rangle } \right)} + a \sum\limits_{i = 0}^{N - 1} 
{^{'}\left( {\left| i \right\rangle \otimes \left| 1 \right\rangle } \right)} \\ 
 \,\,\,\,\,\,\,\,\,\,\,\,\,\,\,\,\,\,\,\,\,\,\,\,\,\,\,\,\,\,\,\,\,\,\,\,\,\,\,\,\,\,\,\,\,\,+ b \sum\limits_{i = 0}^{N - 1} {^{''}\left( {\left| i \right\rangle 
\otimes \left| 0 \right\rangle } \right)} + b \sum\limits_{i = 0}^{N - 
1} {^{''}\left( {\left| i \right\rangle \otimes \left| 1 \right\rangle } 
\right)}. \\ 
 \end{array}
\end{equation}

\noindent
such that,

\begin{equation}
\label{SATeqn21}
Ma^2+(P - M)b^2   = 1.
\end{equation}

Notice that, if no matches exist within the superposition (i.e. $M=0$), then all the amplitudes 
will have positive sign and then applying the diffusion operator $D$ will not change the 
amplitudes of the states as follows: Substituting $\alpha _k = \frac{1}{\sqrt P }$ and 
$\left\langle \alpha \right\rangle = \frac{1}{P}\left( {P\left( 
{\frac{1}{\sqrt P }} \right)} \right)$ in Eqn.(\ref{SATeqn17}) we get:

\begin{equation}
\label{SATeqn22}
\frac{1}{\sqrt P } + \frac{2}{P}\left( {P\left( {\frac{1}{\sqrt P 
}} \right)} \right) = \frac{1}{\sqrt P} = \alpha _k,
\end{equation}


\item[6-] {\it Measurement}. Measure the first $n$ qubits, we get the desired solution with 
probability given below:

\begin{itemize}
\item[i-] Probability $P_{s}$ to find a match out of the $M$ possible matches; taking 
into account that a solution $\left| i \right\rangle $ occurs {\it twice} as: 
$\left( {\left| i \right\rangle \otimes \left| 0 \right\rangle } 
\right)$ with amplitude $b$ and $\left( {\left| i \right\rangle \otimes 
\left| 1 \right\rangle } \right)$ with amplitude $a$ as shown in 
Eqn.(\ref{SATeqn20}), can be calculated as follows:

\begin{equation}
\label{SATeqn24}
\begin{array}{l}
P_s = M(a^2 + b^2)\\
\,\,\,\,\,\,\,\,  = \frac{M}{{2N}}\left( {10 - 16\left( {\frac{M}{N}} \right) + 8\left( {\frac{M}{N}} \right)^2 } \right) \\ 
\,\,\,\,\,\,\,\,  = 5\left( {\frac{M}{N}} \right) - 8\left( {\frac{M}{N}} \right)^2  + 4\left( {\frac{M}{N}} \right)^3.  \\ 
\end{array}
\end{equation}

\item[ii-] Probability $P_{ns}$ to find undesired result out of the states can be 
calculated as follows:

\begin{equation}
\label{SATeqn25}
P_{ns} = (P - 2M)b^2. 
\end{equation}
\end{itemize}

Notice that, using Eqn.(\ref{SATeqn21})

\begin{equation}
\label{SATeqn26}
\begin{array}{l}
 P_s + P_{ns} = M(a^2 + b^2 ) + (P - 2M)b^2 \\ 
 \,\,\,\,\,\,\,\,\,\,\,\,\,\,\,\,\, = Ma^2 + (P - M)b^2 = 1. \\ 
 \end{array}
\end{equation}

\end{itemize}

\subsubsection{Performance after Iterating the Algorithm Once}

\begin{table}[H]
\begin{center}
\begin{tabular}
{|c|c|c|c|}
\hline
 $n$, where $N=2^n$ & 
\ Max. prob.  & 
 Min. prob.  & 
 Avg. prob.   \\ \hline
2  & 
1.0  & 
0.8125  & 
0.875  \\ \hline
 3  & 
 1.0  & 
 0.507812  & 
 0.937500  \\ \hline
 4  & 
 1.0  & 
 0.282227  & 
0.968750 \\ \hline
 5  & 
 1.0  & 
 0.148560  & 
 0.984375 \\ \hline
 6  & 
 1.0  & 
 0.076187  & 
 0.992187  \\ \hline

\end{tabular}
\caption {Algorithm performance with different size search space.}
\label{SATtab3}
\end{center}
\end{table}

Considering Eqn.(\ref{SATeqn18}), Eqn.(\ref{SATeqn19}), Eqn.({\ref{SATeqn24}}) and Eqn.({\ref{SATeqn25}}), we can see that 
the probability to find a solution varies according to the number of matches $M$ in the superposition. 

From Table.\ref{SATtab3}, we can see that the maximum probability is always 1.0,
and the minimum probability (worst case) decreases as the size of the list increases, 
which is expected for small $M$ because the number of states will increase and 
the probability shall distribute over more states while the average 
probability increases as the size of the list increases. It implies 
that the average performance of the algorithm to find a solution increases as the 
size of the list increases.

To verify these results, taking into account that the oracle $U_f$ is taken as a black box, we can define 
the average probability of success of the algorithm; $average(P_s)$, as follows:

\begin{equation}
\label{SATeqn28}
\begin{array}{l}
 average(P_s ) = \frac{1}{{2^N }}\sum\limits_{M = 1}^N {{}^NC_M P_s }  \\ 
 \,\,\,\,\,\,\,\,\,\,\,\,\,\,\,\,\,\,\,\,\,\,\,\,\,\, = \frac{1}{{2^N }}\sum\limits_{M = 1}^N {\frac{{N!}}{{M!(N - M)!}}.M\left( {a^2  + b^2 } \right)}  \\ 
 \,\,\,\,\,\,\,\,\,\,\,\,\,\,\,\,\,\,\,\,\,\,\,\,\,\, = \frac{1}{{2^{N + 1} N^3 }}\sum\limits_{M = 1}^N {\frac{{N!}}{{(M - 1)!(N - M)!}}\left( {10N^2  - 16MN + 8M^2 } \right)}  \\ 
 \,\,\,\,\,\,\,\,\,\,\,\,\,\,\,\,\,\,\,\,\,\,\,\,\,\, = 1-\frac{1}{{2N}}.\\
 \end{array}
\end{equation}

\noindent
where ${}^NC_M  = \frac{{N!}}{{M!(N - M)!}}$ is the number of possible cases for $M$ matches. 


\noindent
We can see that as the size of the list increases $(N \to \infty)$, $average (P_s)$ shown in 
Eqn.(\ref{SATeqn28}) tends to $1$. 


Classically, we can try to find a random guess of the item, 
which represents the solution (one trial guess), we may succeed to find a solution with probability 
$P^{(classical)}_{s} = M/N$. The average probability can be calculated as follows:

\begin{equation}
\label{SATeqn31}
\begin{array}{l}
average(P_s^{(classical)} ) = \frac{1}{{2^N }}\sum\limits_{M = 1}^N {{}^NC_M P_s^{(classical)} } \\
\,\,\,\,\,\,\,\,\,\,\,\,\,\,\,\,\,\,\,\,\,\,\,\,\,\,\,\,\,\,\,\,\,\,\,\,\,\,\,\,\,\,\,\,\,\,\,\,\,\,\,\,\,\,\,= \frac{1}{{2^N }}\sum\limits_{M = 1}^N {\frac{{N.M}}{{M!(N - M)!N}}} \\
\,\,\,\,\,\,\,\,\,\,\,\,\,\,\,\,\,\,\,\,\,\,\,\,\,\,\,\,\,\,\,\,\,\,\,\,\,\,\,\,\,\,\,\,\,\,\,\,\,\,\,\,\,\,\,= \frac{1}{2}. \\ 
\end{array}
\end{equation}

It means that we have an average probability one-half to find or not to find a solution by a single random guess even 
with the increase in the number of matches.

Similarly, Grover's algorithm has an average probabilty {\it one-half} after arbitrary number of iterations 
as we will see. It was shown in \cite{boyer96} that the probability of success of Grover's algorithm after $q$ 
iterations is given by:

\begin{equation}
\label{SATeqn32}
 P^{G^{(q)}}_{s} = \sin ^2 ((2q + 1)\theta ), \,\,\, \mbox{where, }\,0 < \theta  < \frac{\pi }{2}\mbox{ and }\sin ^2 (\theta ) = \frac{M}{N}.\\ 
\end{equation}

The average probability of success of Grover's algorithm after arbitrary number of iterations 
can be calculated as follows (Appendix A):

\begin{equation}
\label{SATeqn33}
 average(P^{G^{(q)}}_{s} ) = \frac{1}{{2^N }}\sum\limits_{M = 1}^N {{}^NC_M \sin ^2 ((2t + 1)\theta )}  = \frac{1}{2}. \\ 
\end{equation}

\begin{figure}[htbp]
\centerline{\includegraphics[width=4.02in,height=3.403in]{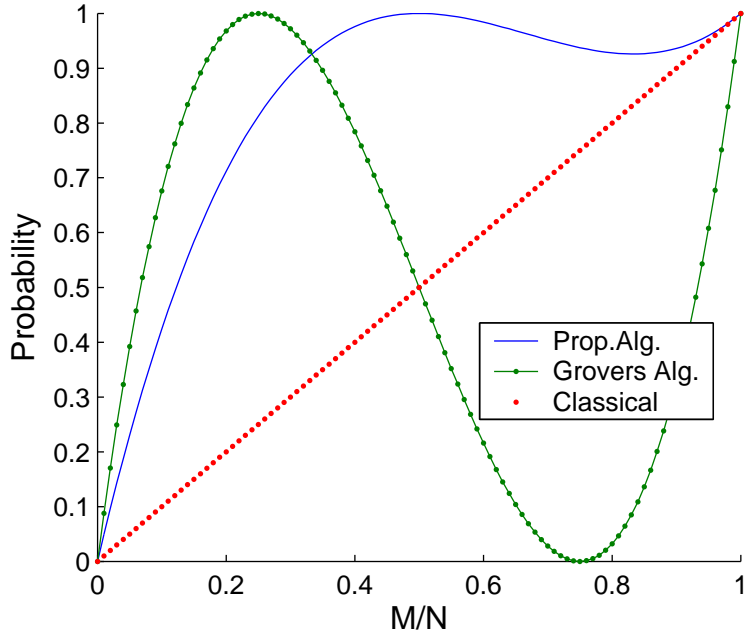}}
\caption{A plot of the probability of success of the proposed algorithm $P_s$, first iteration of Grover's algorithm 
$P^{G^{(1)}}_{s}$ and the classical guess $P_s^{(classical)}$ as a function of the ratio $(M/N)$.} 
\label{SATplot}
\end{figure}

Comparing the performance of the proposed algorithm, first iteration of Grover's algorithm and the classical guess 
technique, Fig.(\ref{SATplot}) shows the probability of success of the three algorithms just mentioned as a function of 
the ratio $(M/N)$.


We can see from Fig.(\ref{SATplot}) that the probability of success of the proposed quantum algorithm is always above 
that of the classical guess technique. Grover's algorithm solves the case where $M=N/4$ with certainty and the proposed 
algorithm solves the case where $M=N/2$ with certainty. The probability of success of Grover's algorithm 
will start to go below one-half for $M>N/2$ while the probability of success of the proposed algorithm will stay 
more reliable with propabilty at least $92.6\%$. For $M<N/8$, the probability of success of the proposed 
algorithm will start to go below one-half where performance of Grover's algorithm will be much better as we will verify 
in the next section.

\subsection{Iterating the algorithm}

\begin{center}
\begin{figure} 
\begin{center}

\setlength{\unitlength}{3947sp}%
\begingroup\makeatletter\ifx\SetFigFont\undefined%
\gdef\SetFigFont#1#2#3#4#5{%
  \reset@font\fontsize{#1}{#2pt}%
  \fontfamily{#3}\fontseries{#4}\fontshape{#5}%
  \selectfont}%
\fi\endgroup%
\begin{picture}(5925,2424)(-449,-2623)
{\color[rgb]{0,0,0}\thinlines
\put(1276,-1636){\circle{150}}
}%
{\color[rgb]{0,0,0}\put(2476,-1861){\circle{150}}
}%
{\color[rgb]{0,0,0}\put(4126,-2461){\circle{150}}
}%
{\color[rgb]{0,0,0}\put(376,-211){\line( 0,-1){1125}}
}%
{\color[rgb]{0,0,0}\put(451,-211){\line(-1, 0){ 75}}
}%
{\color[rgb]{0,0,0}\put(451,-1336){\line(-1, 0){ 75}}
}%
{\color[rgb]{0,0,0}\put(376,-1486){\line( 0,-1){1125}}
}%
{\color[rgb]{0,0,0}\put(451,-1486){\line(-1, 0){ 75}}
}%
{\color[rgb]{0,0,0}\put(451,-2611){\line(-1, 0){ 75}}
}%
{\color[rgb]{0,0,0}\put(451,-361){\line( 1, 0){225}}
}%
{\color[rgb]{0,0,0}\put(676,-436){\framebox(225,225){}}
}%
{\color[rgb]{0,0,0}\put(901,-361){\line( 1, 0){225}}
}%
{\color[rgb]{0,0,0}\put(451,-661){\line( 1, 0){225}}
}%
{\color[rgb]{0,0,0}\put(676,-736){\framebox(225,225){}}
}%
{\color[rgb]{0,0,0}\put(901,-661){\line( 1, 0){225}}
}%
{\color[rgb]{0,0,0}\put(451,-1261){\line( 1, 0){225}}
}%
{\color[rgb]{0,0,0}\put(676,-1336){\framebox(225,225){}}
}%
{\color[rgb]{0,0,0}\put(901,-1261){\line( 1, 0){225}}
}%
{\color[rgb]{0,0,0}\put(451,-1636){\line( 1, 0){975}}
}%
{\color[rgb]{0,0,0}\put(451,-1861){\line( 1, 0){2175}}
}%
{\color[rgb]{0,0,0}\put(451,-2461){\line( 1, 0){3075}}
}%
{\color[rgb]{0,0,0}\put(1276,-1711){\line( 0, 1){375}}
}%
{\color[rgb]{0,0,0}\put(1126,-1336){\framebox(300,1125){}}
}%
{\color[rgb]{0,0,0}\put(1426,-1261){\line( 1, 0){375}}
}%
{\color[rgb]{0,0,0}\put(1426,-661){\line( 1, 0){375}}
}%
{\color[rgb]{0,0,0}\put(1426,-361){\line( 1, 0){375}}
}%
{\color[rgb]{0,0,0}\put(1426,-1711){\framebox(225,225){}}
}%
{\color[rgb]{0,0,0}\put(1651,-1636){\line( 1, 0){150}}
}%
{\color[rgb]{0,0,0}\put(1801,-1711){\framebox(375,1500){}}
}%
{\color[rgb]{0,0,0}\put(2176,-361){\line( 1, 0){150}}
}%
{\color[rgb]{0,0,0}\put(2176,-661){\line( 1, 0){150}}
}%
{\color[rgb]{0,0,0}\put(2176,-1261){\line( 1, 0){150}}
}%
{\color[rgb]{0,0,0}\put(2176,-1636){\line( 1, 0){750}}
}%
{\color[rgb]{0,0,0}\put(2326,-1336){\framebox(300,1125){}}
}%
{\color[rgb]{0,0,0}\put(2476,-1336){\line( 0,-1){600}}
}%
{\color[rgb]{0,0,0}\put(2626,-1936){\framebox(225,225){}}
}%
{\color[rgb]{0,0,0}\put(2851,-1861){\line( 1, 0){ 75}}
}%
{\color[rgb]{0,0,0}\put(2926,-1936){\framebox(375,1725){}}
}%
{\color[rgb]{0,0,0}\put(2626,-1261){\line( 1, 0){300}}
}%
{\color[rgb]{0,0,0}\put(2626,-661){\line( 1, 0){300}}
}%
{\color[rgb]{0,0,0}\put(2626,-361){\line( 1, 0){300}}
}%
{\color[rgb]{0,0,0}\put(3301,-361){\line( 1, 0){225}}
}%
{\color[rgb]{0,0,0}\put(3301,-661){\line( 1, 0){225}}
}%
{\color[rgb]{0,0,0}\put(3301,-1261){\line( 1, 0){225}}
}%
{\color[rgb]{0,0,0}\put(3301,-1636){\line( 1, 0){225}}
}%
{\color[rgb]{0,0,0}\put(3301,-1861){\line( 1, 0){225}}
}%
{\color[rgb]{0,0,0}\put(3751,-361){\line( 1, 0){225}}}%
{\color[rgb]{0,0,0}\put(3751,-661){\line( 1, 0){225}}
}%
{\color[rgb]{0,0,0}\put(3751,-1261){\line( 1, 0){225}}
}%
{\color[rgb]{0,0,0}\put(3976,-1336){\framebox(300,1125){}}
}%
{\color[rgb]{0,0,0}\put(4126,-1936){\line( 0, 1){600}}
}%
{\color[rgb]{0,0,0}\put(4126,-2236){\line( 0,-1){300}}
}%
{\color[rgb]{0,0,0}\put(3751,-2461){\line( 1, 0){525}}
}%
{\color[rgb]{0,0,0}\put(4276,-2536){\framebox(225,225){}}
}%
{\color[rgb]{0,0,0}\put(4501,-2461){\line( 1, 0){150}}
}%
{\color[rgb]{0,0,0}\put(3751,-1861){\line( 1, 0){900}}
}%
{\color[rgb]{0,0,0}\put(3751,-1636){\line( 1, 0){900}}
}%
{\color[rgb]{0,0,0}\put(4276,-1261){\line( 1, 0){375}}
}%
{\color[rgb]{0,0,0}\put(4276,-661){\line( 1, 0){375}}
}%
{\color[rgb]{0,0,0}\put(4276,-361){\line( 1, 0){375}}
}%
{\color[rgb]{0,0,0}\put(4651,-2536){\framebox(375,2325){}}
}%
{\color[rgb]{0,0,0}\put(5026,-361){\line( 1, 0){225}}
}%
{\color[rgb]{0,0,0}\put(5026,-661){\line( 1, 0){225}}
}%
{\color[rgb]{0,0,0}\put(5026,-1261){\line( 1, 0){225}}
}%
{\color[rgb]{0,0,0}\put(5026,-1636){\line( 1, 0){225}}
}%
{\color[rgb]{0,0,0}\put(5026,-1861){\line( 1, 0){225}}
}%
{\color[rgb]{0,0,0}\put(5026,-2461){\line( 1, 0){225}}
}%
{\color[rgb]{0,0,0}\put(5401,-211){\line( 0,-1){1125}}
}%
{\color[rgb]{0,0,0}\put(5401,-211){\line(-1, 0){ 75}}
}%
{\color[rgb]{0,0,0}\put(5401,-1336){\line(-1, 0){ 75}}
}%
{\color[rgb]{0,0,0}\put(5176,-1000){$\vdots$}}%
{\color[rgb]{0,0,0}\put(526,-1000){$\vdots$}}%
{\color[rgb]{0,0,0}\put(770,-1000){$\vdots$}}%
{\color[rgb]{0,0,0}\put(5176,-2180){$\vdots$}}%
{\color[rgb]{0,0,0}\put(4120,-2180){$\vdots$}}%
{\color[rgb]{0,0,0}\put(526,-2180){$\vdots$}}%
{\color[rgb]{0,0,0}\put(3550,-377){$\ldots$}}%
{\color[rgb]{0,0,0}\put(3550,-677){$\ldots$}}%
{\color[rgb]{0,0,0}\put(3550,-1277){$\ldots$}}%
{\color[rgb]{0,0,0}\put(3550,-1657){$\ldots$}}%
{\color[rgb]{0,0,0}\put(3550,-1867){$\ldots$}}%
{\color[rgb]{0,0,0}\put(3550,-2477){$\ldots$}}%
\put(-74,-661){$n$}%
\put(-224,-886){qubits}%
\put(-74,-1861){$q$}%
\put(-224,-2086){qubits}%
\put(-490,-2311){workspace}%
\put(720,-361){$H$}%
\put(720,-661){$H$}%
\put(720,-1261){$H$}%
\put(1470,-1636){$H$}%
\put(4320,-2461){$H$}%
\put(1201,-886){$U_f$}%
\put(2401,-886){$U_f$}%
\put(4051,-886){$U_f$}%
\put(2670,-1861){$H$}%
\put(1900,-886){$D$}%
\put(3030,-886){$D$}%
\put(4780,-886){$D$}%
\put(5476,-886){Measure}%
\end{picture}
\end{center}
\caption{Quantum circuit for the iterative version of the proposed algorithm.}
\label{SATitr}
\end{figure}
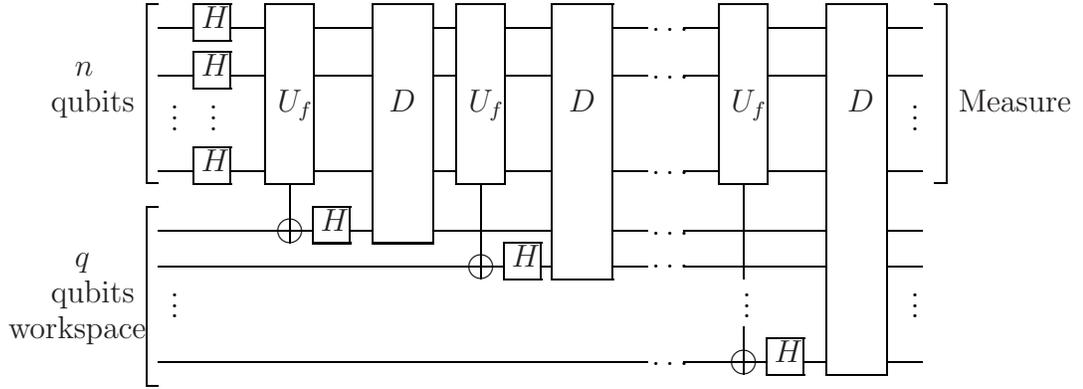
\end{center}

If we consider iterating the above algorithm: For a list of size $N(=2^n)$, 
prepare $n$ qubits and append extra $q$ qubits for applying $q$ 
iterations of the algorithm. The iterating version of the algorithm works as follows 
(as shown in Fig.(\ref{SATitr})):

\begin{itemize}

\item[1-]	Initialize the whole $n+q$ qubits system to the state $\left| 0 \right\rangle $.
\item[2-]	Apply Hadamard gate on each of the first $n$ qubits in parallel.
\item[3-]	Iterate the following, for iteration $k$:
\begin{itemize}
\item[a.]	Apply the oracle $U_{f}$ taking the first $n$ qubits as control qubits and the 
$k^{th}$ qubit workspace as the target qubit exclusively.

\item[b.]	Apply Hadamard gate on the $k^{th}$ qubit workspace.

\item[c.]	Apply diffusion operator on the whole $n+k$ qubit system inclusively.
\end{itemize}
\item[4-]	Apply measurement on the first $n$ qubits.
\end{itemize}

To understand how the iterative version of the algorithm affects the system, 
we will trace the state of the system during the first few iterations.

Consider the system after the first iteration shown in Eqn.(\ref{SATeqn20}), second 
iteration will modify the system as follows (to clear ambiguity, $a$ and $b$ used 
in the above section will be denoted as $a_0^{(1)} $ and $b_0^{(1)} $ 
respectively, where the {\it superscript index} denotes the iteration 
and the {\it subscript index} is used to distinguish amplitudes):

\begin{itemize}
\item[1-] Append second qubit workspace to the system:

\begin{equation}
\label{SATeqn34}
\begin{array}{l}
 \left| {W_1^{(2)}} \right\rangle = b_0^{(1)} \sum\limits_{i = 0}^{N - 1} 
{'\left( {\left| i \right\rangle \otimes \left| 0 \right\rangle } \right) 
\otimes \left| 0 \right\rangle } + a_0^{(1)} \sum\limits_{i = 0}^{N - 1} 
{'\left( {\left| i \right\rangle \otimes \left| 1 \right\rangle } \right) 
\otimes \left| 0 \right\rangle } \\ 
    \,\,\,\,\,\,\,\,\,\,\,\,\,\,\,\,\,\,\,\,\,\,\,\, + b_0^{(1)} \sum\limits_{i = 0}^{N - 1} 
{''\left( {\left| i \right\rangle \otimes \left| 0 \right\rangle } \right) 
\otimes \left| 0 \right\rangle } + b_0^{(1)} \sum\limits_{i = 0}^{N - 1} 
{''\left( {\left| i \right\rangle \otimes \left| 1 \right\rangle } \right) 
\otimes \left| 0 \right\rangle }. \\ 
\end{array}
\end{equation}

\item[2-]Apply $U_{f}$ as shown in step 3-a:

\begin{equation}
\label{SATeqn35}
\begin{array}{l}
 \left| {W_2^{(2)} } \right\rangle = b_0^{(1)} \sum\limits_{i = 0}^{N - 1} 
{'\left( {\left| i \right\rangle \otimes \left| 0 \right\rangle } \right) 
\otimes \left| 1 \right\rangle } + a_0^{(1)} \sum\limits_{i = 0}^{N - 1} 
{'\left( {\left| i \right\rangle \otimes \left| 1 \right\rangle } \right) 
\otimes \left| 1 \right\rangle } \\ 
 \,\,\,\,\,\,\,\,\,\,\,\,\,\,\,\,\,\,\,\,\,\,\,\, + b_0^{(1)} \sum\limits_{i = 0}^{N - 1} 
{''\left( {\left| i \right\rangle \otimes \left| 0 \right\rangle } \right) 
\otimes \left| 0 \right\rangle } + b_0^{(1)} \sum\limits_{i = 0}^{N - 1} 
{''\left( {\left| i \right\rangle \otimes \left| 1 \right\rangle } \right) 
\otimes \left| 0 \right\rangle }. \\ 
 \end{array}
\end{equation}

\item[3-]Apply Hadamard gate on second qubit workspace $\left( {I^{ \otimes n + 1} 
\otimes H} \right)$:

\begin{equation}
\label{SATeqn36}
\begin{array}{l}
 \left| {W_3^{(2)} } \right\rangle = \frac{b_0^{(1)} }{\sqrt 2 }\sum\limits_{i = 
0}^{N - 1} {'\left( {\left| i \right\rangle \otimes \left| 0 \right\rangle } 
\right) \otimes \left| 0 \right\rangle } - \frac{b_0^{(1)} }{\sqrt 2 
}\sum\limits_{i = 0}^{N - 1} {'\left( {\left| i \right\rangle \otimes \left| 
0 \right\rangle } \right) \otimes \left| 1 \right\rangle } \\ 
 \,\,\,\,\,\,\,\,\,\,\,\,\,\,\,\,\,\,\,\,\,\,\,\, + \frac{a_0^{(1)} }{\sqrt 2 }\sum\limits_{i 
= 0}^{N - 1} {'\left( {\left| i \right\rangle \otimes \left| 1 \right\rangle 
} \right) \otimes \left| 0 \right\rangle - \,} \frac{a_0^{(1)} }{\sqrt 2 
}\sum\limits_{i = 0}^{N - 1} {'\left( {\left| i \right\rangle \otimes \left| 
1 \right\rangle } \right) \otimes \left| 1 \right\rangle } \\ 
 \,\,\,\,\,\,\,\,\,\,\,\,\,\,\,\,\,\,\,\,\,\,\,\, + \frac{b_0^{(1)} }{\sqrt 2 }\sum\limits_{i 
= 0}^{N - 1} {''\left( {\left| i \right\rangle \otimes \left| 0 
\right\rangle } \right) \otimes \left| 0 \right\rangle } + \frac{b_0^{(1)} 
}{\sqrt 2 }\sum\limits_{i = 0}^{N - 1} {''\left( {\left| i \right\rangle 
\otimes \left| 0 \right\rangle } \right) \otimes \left| 1 \right\rangle } \\ 
 \,\,\,\,\,\,\,\,\,\,\,\,\,\,\,\,\,\,\,\,\,\,\,\, + \frac{b_0^{(1)} }{\sqrt 2 }\sum\limits_{i = 
0}^{N - 1} {''\left( {\left| i \right\rangle \otimes \left| 1 \right\rangle 
} \right) \otimes \left| 0 \right\rangle } + \frac{b_0^{(1)} }{\sqrt 2 
}\sum\limits_{i = 0}^{N - 1} {''\left( {\left| i \right\rangle \otimes 
\left| 1 \right\rangle } \right) \otimes \left| 1 \right\rangle }.\\ 
 \end{array}
\end{equation}

\item[4-]Apply diffusion operator as shown in step 3-c:

\begin{equation}
\label{SATeqn37}
\begin{array}{l}
 \left| {W_4^{(2)} } \right\rangle = b_0^{(2)} \sum\limits_{i = 0}^{N - 1} 
{'\left( {\left| i \right\rangle \otimes \left| 0 \right\rangle } \right) 
\otimes \left| 0 \right\rangle } + b_1^{(2)} \sum\limits_{i = 0}^{N - 1} 
{'\left( {\left| i \right\rangle \otimes \left| 0 \right\rangle } \right) 
\otimes \left| 1 \right\rangle } \\ 
 \,\,\,\,\,\,\,\,\,\,\,\,\,\,\,\,\,\,\,\,\,\,\,\, + a_0^{(2)} \sum\limits_{i = 0}^{N - 1} 
{'\left( {\left| i \right\rangle \otimes \left| 1 \right\rangle } \right) 
\otimes \left| 0 \right\rangle + \,} a_1^{(2)} \sum\limits_{i = 0}^{N - 1} 
{'\left( {\left| i \right\rangle \otimes \left| 1 \right\rangle } \right) 
\otimes \left| 1 \right\rangle } \\ 
 \,\,\,\,\,\,\,\,\,\,\,\,\,\,\,\,\,\,\,\,\,\,\,\, + b_0^{(2)} \sum\limits_{i = 0}^{N - 1} 
{''\left( {\left| i \right\rangle \otimes \left| 0 \right\rangle } \right) 
\otimes \left| 0 \right\rangle } + b_0^{(2)} \sum\limits_{i = 0}^{N - 1} 
{''\left( {\left| i \right\rangle \otimes \left| 0 \right\rangle } \right) 
\otimes \left| 1 \right\rangle } \\ 
 \,\,\,\,\,\,\,\,\,\,\,\,\,\,\,\,\,\,\,\,\,\,\,\, + b_0^{(2)} \sum\limits_{i = 0}^{N - 1} 
{''\left( {\left| i \right\rangle \otimes \left| 1 \right\rangle } \right) 
\otimes \left| 0 \right\rangle } + b_0^{(2)} \sum\limits_{i = 0}^{N - 1} 
{''\left( {\left| i \right\rangle \otimes \left| 1 \right\rangle } \right) 
\otimes \left| 1 \right\rangle }. \\ 
 \end{array}
\end{equation}

\noindent
where the mean of the amplitudes to be used in the diffusion operator is 
calculated as follows:

\begin{equation}
\label{SATeqn38}
\begin{array}{l}
 \left\langle {\alpha _2 } \right\rangle = \frac{1}{2^{n + 2}}\left( {\left( 
{2^{n + 2} - 4M} \right)\frac{b_0^{(1)} }{\sqrt 2 }} \right) \\ 
 \,\,\,\,\,\,\,\,\,\,\,\, = \frac{b_0^{(1)} }{\sqrt 2 }\left( {1 - \frac{M}{N}} 
\right). \\ 
 \end{array}
\end{equation}

\noindent 
And the new amplitudes $a_0^{(2)} $, $a_1^{(2)} $,$b_0^{(2)} $ and 
$b_1^{(2)} $ are calculated as follows:

\begin{equation}
\label{SATeqn39}
\begin{array}{l}
 a_0^{(2)} = 2\left\langle {\alpha _2 } \right\rangle - \frac{a_0^{(1)} 
}{\sqrt 2 };\,\,\,\,a_1^{(2)} = 2\left\langle {\alpha _2 } \right\rangle + 
\frac{a_0^{(1)} }{\sqrt 2 }. \\ 
 b_0^{(2)} = 2\left\langle {\alpha _2 } \right\rangle - \frac{b_0^{(1)} 
}{\sqrt 2 };\,\,\,\,\,b_1^{(2)} = 2\left\langle {\alpha _2 } \right\rangle + 
\frac{b_0^{(1)} }{\sqrt 2 }. \\ 
\end{array}
\end{equation}

And the probability of success:

\begin{equation}
\label{SATeqn40}
P_s^{\left( 2 \right)} = M\left( {\left( {a_0^{(2)} } \right)^2 + \left( 
{a_1^{(2)} } \right)^2 + \left( {b_0^{(2)} } \right)^2 + \left( {b_1^{(2)} } 
\right)^2} \right).
\end{equation}

\end{itemize}

For the sake of simplicity, we can trace the effect of each iteration on 
the amplitudes of the system instead of writing the state of the system 
explicitly; for example, the amplitudes of the system after third iteration 
will be as follows:

\begin{itemize}

\item[1-] The mean of the amplitudes to be used in the diffusion operator: 

\begin{equation}
\label{SATeqn41}
\begin{array}{l}
 \left\langle {\alpha _3 } \right\rangle = \frac{1}{2^{n + 3}}\left( {\left( 
{2^{n + 3} - 8M} \right)\frac{b_0^{(2)} }{\sqrt 2 }} \right) \\ 
 \,\,\,\,\,\,\,\,\,\,\,\, = \frac{b_0^{(2)} }{\sqrt 2 }\left( {1 - \frac{M}{N}} 
\right). \\ 
 \end{array}
\end{equation}

\item[2-]The new amplitudes:

\begin{equation}
\label{SATeqn42}
\begin{array}{l}
 a_0^{(3)} = 2\left\langle {\alpha _3 } \right\rangle - \frac{a_0^{(2)} 
}{\sqrt 2 };\,\,\,\,a_1^{(3)} = 2\left\langle {\alpha _3 } \right\rangle + 
\frac{a_0^{(2)} }{\sqrt 2 }. \\ 
 a_2^{(3)} = 2\left\langle {\alpha _3 } \right\rangle - \frac{a_1^{(2)} 
}{\sqrt 2 };\,\,\,\,a_3^{(3)} = 2\left\langle {\alpha _3 } \right\rangle + 
\frac{a_1^{(2)} }{\sqrt 2 }. \\ 
 b_0^{(3)} = 2\left\langle {\alpha _3 } \right\rangle - \frac{b_0^{(2)} 
}{\sqrt 2 };\,\,\,\,b_1^{(3)} = 2\left\langle {\alpha _3 } \right\rangle + 
\frac{b_0^{(2)} }{\sqrt 2 }. \\ 
 b_2^{(3)} = 2\left\langle {\alpha _3 } \right\rangle - \frac{b_1^{(2)} 
}{\sqrt 2 };\,\,\,\,b_3^{(3)} = 2\left\langle {\alpha _3 } \right\rangle + 
\frac{b_1^{(2)} }{\sqrt 2 }. \\ 
 \end{array}
\end{equation}

\item[3-] And the probability of success:

\begin{equation}
\label{SATeqn43}
P_s^{\left( 3 \right)} = M\left( {\left( {a_i^{(3)} } \right)^2 + \left( 
{b_i^{(3)} } \right)^2} \right);i = 0,1,2,3.
\end{equation}

\end{itemize}
In general, after $q$ iterations the recurrence relations representing the iteration can 
be written as follows:

The initial conditions: $a_0^{(0)}  = b_0^{(0)}  = \frac{1}{{\sqrt N }}$.

\begin{itemize}

\item[1-] The mean to be used in the diffusion operator: 

\begin{equation}
\label{SATeqn44}
\left\langle {\alpha _q } \right\rangle \, = \frac{b_0^{(q - 1)} }{\sqrt 2 
}\left( {1 - \frac{M}{N}} \right); q\ge1.
\end{equation}

\item[2-] The new amplitudes of the system:

\begin{equation}
\label{SATeqn45}
a_0^{(1)}  = 2\left\langle {\alpha _1 } \right\rangle  + \frac{{a_0^{(0)} }}{{\sqrt 2 }},\,\,\,\,
a_{0 \to 2^{q - 1} - 1}^{(q)} = 2\left\langle {\alpha _q } \right\rangle \mp \frac{a_{0 \to 2^{q - 2} - 1}^{(q - 1)} }{\sqrt 2 };\,\,\,\,q\ge2.
\end{equation}

\begin{equation}
\label{SATeqn46}
b_0^{(1)}  = 2\left\langle {\alpha _1 } \right\rangle  - \frac{{b_0^{(0)} }}{{\sqrt 2 }},\,\,\,\,b_{0 \to 2^{q - 1} - 1}^{(q)} = 2\left\langle {\alpha _q } \right\rangle \mp 
\frac{b_{0 \to 2^{q - 2} - 1}^{(q - 1)} }{\sqrt 2 };\,\,\,\,q\ge2.
\end{equation}

\item[3-] The probability of success for $q\ge1$:

\begin{equation}
\label{SATeqn47}
P_s^{\left( q \right)} = M\left( {\left( {a_i^{(q)} } \right)^2 + \left( 
{b_i^{(q)} } \right)^2} \right);i = 0,1,2,...,2^{q - 1} - 1.
\end{equation}

\end{itemize}

Using mathematical induction, we can prove that the probability of success 
after $q$ iterations shown in Eqn.(\ref{SATeqn47}) can take this form (Appendix B):

\begin{equation}
\label{SATeqn48}
P_s^{(q)} = \left( {\frac{M}{N} - 1} \right)\left( {1 - \frac{2M}{N}} 
\right)^{2q} + 1,\,\,\,\,\,\, q\ge1.
\end{equation}

\subsubsection{Performance of Iterating the Algorithm}

\begin{itemize}

\item[i-] The case where multiple instances of a match exist within the search 
space: Consider the following cases using Eqn.(\ref{SATeqn48}):

\begin{itemize}

\item[1-]The case where $M = N / 2$: the algorithm can find a solution with 
{\it certainty} after arbitrary number of iterations (one iteration is enough). 

\item[2-]The case where $M > N / 2$: the probability of success is; for instances, at least 
92.6{\%} after the first iteration, 95.9{\%} after second iteration and  97.2{\%} after third iteration.

\item[3-]For iterating the algorithm once ($q = 1$) and to get probability at least 
one-half, so, $M$ must satisfy the condition $M \ge N / 8$.

\end{itemize}

\item[ii-] The case where few instances of a match exist within the search space:

First, we need to represent the number of iterations $q$ in terms of the ratio $M/N$. 
From Eqn.(\ref{SATeqn48}) and using Taylor's expansion we get:

\begin{equation}
\label{SATeqn49}
q \ge \frac{P_s^{(q)} - \textstyle{M \over N}}{4\textstyle{M \over N}\left( 
{1 - \textstyle{M \over N}} \right)}.
\end{equation}

For the cases where $q > 1$, the following conditions must be satisfied:

\begin{equation}
\label{SATeqn50}
n \ge 4 \mbox{ and } 1 \le M < N / 8.
\end{equation}

\end{itemize}

\begin{figure}[htbp]
\centerline{\includegraphics[width=3.278in,height=2.874in]{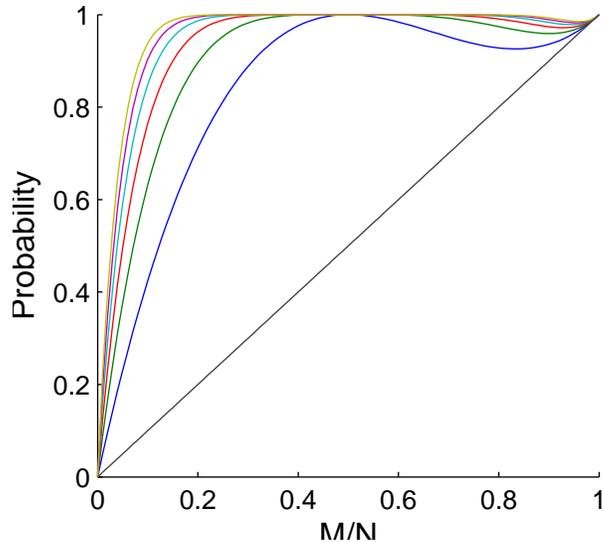}}
\caption{A plot of the probability of success of the iterative version of the 
proposed algorithm where $q$ = 1,2,\ldots ,6.}
\label{6itrs}
\end{figure}

It means that first iteration will cover approximately 87.5{\%} of the problem with 
probability at least one-half; two iterations will cover approximately 92{\%} and three iterations will 
cover 94{\%}. It is easy to prove that the rate of 
the increase of the coverage range will decrease as number of iterations 
increases as shown in Fig.(\ref{6itrs}). We can also see from Eqn.(\ref{SATeqn49}) and Eqn.(\ref{SATeqn50}) that the algorithm 
needs $O\left( {N / M} \right)$ iterations for $n \ge 4$ and $1 \le M < N / 
8$, which is similar to classical algorithms behaviour. It leads to a 
conclusion that first few iterations of the algorithm will do the best 
performance and there will be no big gain from continuing to iterate the algorithm.

\section{A Hybrid Quantum Search Engine}

We have devised a quantum search algorithm, which performs very well in 
case of multiple instances of the solution within the search space and a 
classical behaviour in case of few instances of the solution. On 
contrary, Grover's algorithm needs $O\left( {\sqrt {N / M} } \right)$ to 
solve the problem but it's performance decreases for $M > N / 2$ \cite{niel00}.

This leads up to propose a {\it hybrid quantum search engine,} 
which combines both algorithms and can be integrated as follows:

\begin{itemize}
\item[i-]If the number of solutions $M$ is {\it known in advance}:

\begin{itemize}

\item[1-]If $1 \le M < N / 8$: Use Grover's algorithm with $O\left( {\sqrt {N / M} 
} \right)$.

\item[2-] If $N / 8 \le M < N$: Use the proposed algorithm with $O(1)$ .
\end{itemize}

\item[ii-]If the number of solutions $M$ is {\it unknown}:

Iterate the proposed algorithm few times; say three iterations, which results 
in a chance of approximately 94{\%} to find a solution. If it fails, we apply 
Grover's algorithm so we still have the same complexity $O\left( {\sqrt {N / M} 
} \right)$.
\end{itemize}

\begin{figure}[htbp]
\centerline{\includegraphics[width=3.208in,height=2.431in]{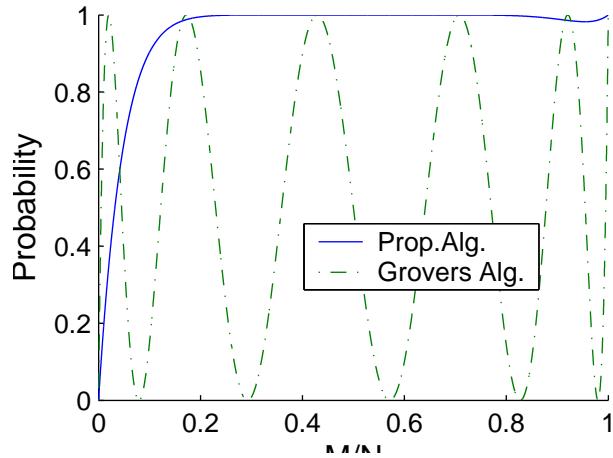}}
\caption{The probability of success after five iterations from Grover's algorithm's 
vs. the proposed algorithm.}
\label{5thitrcomp}
\end{figure}

We can see from Fig.(\ref{5thitrcomp}) that Grover's 
algorithm is much faster in the case of few instances of the solution (ratio 
$M/N$ is small) and the proposed algorithm is more stable and reliable in 
case of multiple instances of the solution.

\section{Conclusion}

In this paper, we proposed a quantum search algorithm, which performs very fast in 
the case of multiple instances of the solution within the search space 
(almost constant run-time) but the performance turns out to be classical for few 
instances of the solution. 


On the other hand we have Grover's algorithm, 
which performs very well in case of few instances of the solution and the 
performance decrease as number of solutions increase within the search 
space.

This gave us the chance to propose a {\it hybrid quantum search engine} with 
{\it general performance} better that any pure classical or 
quantum search algorithm and still has $O\left( {\sqrt N } \right)$ for the 
hardest case and approximately $O(1)$ for $M \ge N / 8$.

\section*{Appendix A}

To proof the identity shown in Eqn.(\ref{SATeqn33}), we first need the next Lemma.

\begin{lemma}
Let $\alpha$ and $\beta$ angles; $0 < \alpha,\beta < \pi/2$, such that :

\begin{equation}
\label{lem1}
\sin ^2 \left( \alpha  \right) = \cos ^2 \left( \beta  \right).
\end{equation}

Then, if $k$ is any odd positive integer, we have:

\begin{equation}
\label{lem2}
\sin ^2 \left( k\alpha  \right) = \cos ^2 \left( k\beta  \right).
\end{equation}

\begin{proof}
Since, $0 < \alpha,\beta < \pi/2$, then we have, 

\begin{equation}
\label{lem3}
\sin \left( \alpha  \right) = \cos \left( \beta  \right).
\end{equation}

Also, we can write the following: $\cos \left( \beta  \right)=\sin \left( \pi/2-\beta \right) = \sin \left( \alpha \right)$.

So, $\alpha=\pi/2-\beta$.

Therefore;

\begin{equation}
\begin{array}{l}
 \cos \left( {k\beta } \right) = \cos \left( {k\left( {\frac{\pi }{2} - \alpha } \right)} \right) \\ 
 \,\,\,\,\,\,\,\,\,\,\,\,\,\,\,\,\,\,\, = \cos \left( {\frac{{k\pi }}{2} - k\alpha } \right) \\ 
 \,\,\,\,\,\,\,\,\,\,\,\,\,\,\,\,\,\,\, = \cos \left( {\frac{{k\pi }}{2}} \right)\cos \left( {k\alpha } \right) + \sin \left( {\frac{{k\pi }}{2}} \right)\sin \left( {k\alpha } \right)\\
 \,\,\,\,\,\,\,\,\,\,\,\,\,\,\,\,\,\,\, =\pm \sin \left( {k\beta } \right).\\
 \end{array}
\end{equation}

Then, $\cos ^2 \left( {k\beta } \right) = \sin ^2 \left( {k\beta } \right).$

\end{proof}
\end{lemma}

\begin{theorem}
For any odd positive integer $k$ and an angle ${\theta _M }$; $0 < \theta _M  < \frac{\pi }{2}$, defined as follows:
\begin{equation}
\sin ^2 \left( {\theta _M } \right) = \frac{M}{N}.
\end{equation}

Then, 
\begin{equation}\sum\limits_{M = 1}^N {{}^NC_M \sin ^2 \left( {k\theta _M } \right)}  = 2^{n - 1}.\end{equation}

\begin{proof}

Consider 
\begin{equation}
sin ^2 \left( {\theta _M } \right) + \sin ^2 \left( {\theta _{N - M} } \right) = \frac{M}{N} + \frac{{N - M}}{N} = 1.
\end{equation}

So, 
\begin{equation}
\begin{array}{l}
\sin ^2 \left( {\theta _M } \right) = 1 - \sin ^2 \left( {\theta _{N - M} } \right)\\
\,\,\,\,\,\,\,\,\,\,\,\,\,\,\,\,\,\,\,\,\,\,\,\,\,\,=\cos ^2 \left( {\theta _{N - M} } \right).
\end{array}
\end{equation}

By Lemma, 
\begin{equation}
\sin ^2 \left( {k\theta _M } \right) = \cos ^2 \left( {k\theta _{N - M} } \right). 
\end{equation}

Or,
 
\begin{equation}\sin ^2 \left( {k\theta _M } \right) + \sin ^2 \left( {k\theta _{N - M} } \right) = 1.\end{equation}

Now consider,
\begin{equation}
\begin{array}{l}
 2\sum\limits_{M = 1}^N {{}^NC_M \sin ^2 \left( {k\theta _M } \right)}  = \sum\limits_{M = 1}^N {{}^NC_M \sin ^2 \left( {k\theta _M } \right)}  + \sum\limits_{M = 1}^N {{}^NC_{N - M} \sin ^2 \left( {k\theta _M } \right)}  \\ 
 \,\,\,\,\,\,\,\,\,\,\,\,\,\,\,\,\,\,\,\,\,\,\,\,\,\,\,\,\,\,\,\,\,\,\,\,\,\,\,\,\,\,\,\, = \sum\limits_{M = 1}^N {{}^NC_M \sin ^2 \left( {k\theta _M } \right)}  + \sum\limits_{M = 0}^N {{}^NC_M \sin ^2 \left( {k\theta _{N - M} } \right)}  \\ 
 \,\,\,\,\,\,\,\,\,\,\,\,\,\,\,\,\,\,\,\,\,\,\,\,\,\,\,\,\,\,\,\,\,\,\,\,\,\,\,\,\,\,\,\, = \sum\limits_{M = 1}^N {{}^NC_M \left( {\sin ^2 \left( {k\theta _M } \right) + \sin ^2 \left( {k\theta _{N - M} } \right)} \right)}  \\ 
 \,\,\,\,\,\,\,\,\,\,\,\,\,\,\,\,\,\,\,\,\,\,\,\,\,\,\,\,\,\,\,\,\,\,\,\,\,\,\,\,\,\,\,\, = \sum\limits_{M = 1}^N {{}^NC_M }  \\ 
 \,\,\,\,\,\,\,\,\,\,\,\,\,\,\,\,\,\,\,\,\,\,\,\,\,\,\,\,\,\,\,\,\,\,\,\,\,\,\,\,\,\,\,\, = 2^n. \\ 
 \end{array}
\end{equation}

\end{proof}
\end{theorem}

\newpage
\section*{Appendix B}

To prove that the probability of success after $q$ iterations is as shown in 
Eqn.(\ref{SATeqn48}), we need first to prove the following relation: 

Let $b_0^{(0)}=\frac{1}{\sqrt {2^n} } =\frac{1}{\sqrt {N} }$, and given by the definition of the 
diffusion operator for $q\ge1$ that,

\begin{equation}
\left\langle {\alpha _q } \right\rangle = \frac{b_0^{(q - 1)} }{\sqrt 2 
}\left( {1 - \frac{M}{N}} \right)
\end{equation}

And,

\begin{equation}
b_0^{(q)} = 2\left\langle {\alpha _q } \right\rangle - \frac{b_0^{(q - 1)} 
}{\sqrt 2 }
\end{equation}

Then,

\begin{equation}
\label{SATappB}
b_0^{(q)} = \frac{b_0^{(0)}}{\left( {\sqrt 2 } \right)^q}\left( {1 - 2\frac{M}{N}} 
\right)^q
\end{equation}

\begin{proof} 
(By Mathematical Induction)

\begin{itemize}

\item[Step 1:] For $q=1$, it follows directly from Eqn.(\ref{SATeqn19}) as follows:

\begin{equation}
\begin{array}{l}
 b_0^{(1)} = \frac{1}{\sqrt {2^{n + 1}} }\left( {1 - 2\frac{M}{N}} \right) 
\\ 
 \,\,\,\,\,\,\,\, = \frac{b_0^{(0)}}{\sqrt 2 }\left( {1 - 2\frac{M}{N}} \right) \\ 
 \end{array}
\end{equation}

\item[Step 2:] Assume the relation is true for $q=t$:

\begin{equation}
b_0^{(t)} = \frac{b_0^{(0)}}{\left( {\sqrt 2 } \right)^t}\left( {1 - 2\frac{M}{N}} 
\right)^t
\end{equation}

\item[Step 3:] Prove for $q=t+1$:

By definition,

\begin{equation}
\left\langle {\alpha _{t + 1} } \right\rangle = \frac{b_0^{(t)} }{\sqrt 2 
}\left( {1 - \frac{M}{N}} \right)
\end{equation}

And,

\begin{equation}
\begin{array}{l}
 b_0^{(t + 1)} = 2\left\langle {\alpha _{t + 1} } \right\rangle - 
\frac{b_0^{(t)} }{\sqrt 2 } \\ 
 \,\,\,\,\,\,\,\,\,\,\, = \frac{2b_0^{(t)} }{\sqrt 2 }\left( {1 - 
\frac{M}{N}} \right) - \frac{b_0^{(t)} }{\sqrt 2 } \\ 
 \,\,\,\,\,\,\,\,\,\,\, = \frac{b_0^{(t)} }{\sqrt 2 }\left( {1 - 
2\frac{M}{N}} \right) \\ 
 \end{array}
\end{equation}

Substitute by the assumption, it directly gives the term for $q=t+1$,

\begin{equation}
b_0^{(t + 1)} = \frac{b_0^{(0)}}{\left( {\sqrt 2 } \right)^{t + 1}}\left( {1 - 
2\frac{M}{N}} \right)^{t + 1}
\end{equation}
\end{itemize}
\end{proof}

Now, to prove that the probability of success of the proposed algorithm 
after $q$ iterations can take this form:

\begin{equation}
P_s^{(q)} = \left( {\frac{M}{N} - 1} \right)\left( {1 - \frac{2M}{N}} 
\right)^{2q} + 1.
\end{equation}

Given by definition that,

\begin{equation}
P_s^{\left( q \right)} = M\left( {\left( {a_i^{(q)} } \right)^2 + \left( 
{b_i^{(q)} } \right)^2} \right);i = 0,1,2,...,2^{q - 1} - 1.
\end{equation}

\begin{proof} 
(By Mathematical Induction)

\begin{itemize}
\item[Step 1:] For $q=1$, it is straight forward from Eqn.(24).

\item[Step 2:] Assume the relation is true for $q=t$,

\begin{equation}
\begin{array}{l}
 P_s^{(t)} = M\left( {\left( {a_i^{(t)} } \right)^2 + \left( {b_i^{(t)} } 
\right)^2} \right);i = 0,1,...,2^{t - 1} - 1. \\ 
 \,\,\,\,\,\,\,\,\,\, = \left( {\frac{M}{N} - 1} \right)\left( {1 - 
\frac{2M}{N}} \right)^{2t} + 1 \\ 
 \end{array}.
\end{equation}

\item[Step 3:] Proof for $q=t+1$,

By definition,

\begin{equation}
\label{SATappB1}
\begin{array}{l}
 P_s^{\left( {t + 1} \right)} = M\left( {\left( {a_{0 \to 2^t - 1}^{(t + 1)} 
} \right)^2 + \left( {b_{0 \to 2^t - 1}^{(t + 1)} } \right)^2} \right) \\ 
= M\left( {2^{t + 2}\left\langle {\alpha _{t + 1} } \right\rangle ^2 + 
\left( {a_{0 \to 2^{t - 1} - 1}^{(t + 1)} } \right)^2 + 2^{t + 
2}\left\langle {\alpha _{t + 1} } \right\rangle ^2 + \left( {b_{0 \to 2^{t - 
1} - 1}^{(t + 1)} } \right)^2} \right) \\  
 = M2^{t + 3}\left\langle {\alpha _{t + 1} } \right\rangle ^2 + P_s^{(t)} \\ 
 \end{array}
\end{equation}

Using Eqn.(\ref{SATappB}), we have,

\begin{equation}
\begin{array}{l}
 \left\langle {\alpha _{t + 1} } \right\rangle ^2 = \left( {\frac{b_0^{(t)} 
}{\sqrt 2 }\left( {1 - \frac{M}{N}} \right)} \right)^2 \\ 
\,\,\,\,\,\,\,\,\,\,\,\,\,\,\,\,\,\,\,\,\, = \left( {\frac{b_0^{(0)}}{\left( {\sqrt 2 } \right)^{t + 1}}} \right)^2\left( {1 - 
\frac{M}{N}} \right)^2\left( {1 - \frac{2M}{N}} \right)^{2t} \\ 
 \,\,\,\,\,\,\,\,\,\,\,\,\,\,\,\,\,\,\,\,\, = \frac{1}{N2^{t + 1}}\left( {1 - \frac{M}{N}} \right)^2\left( {1 - 
\frac{2M}{N}} \right)^{2t} \\ 
 \end{array}
\end{equation}

Substitute in Eqn.(\ref{SATappB1}), we get,

\begin{equation}
\begin{array}{l}
 P_s^{\left( {t + 1} \right)} = M2^{t + 3}\frac{1}{N2^{t + 1}}\left( {1 - 
\frac{M}{N}} \right)^2\left( {1 - \frac{2M}{N}} \right)^{2t} + \left( 
{\frac{M}{N} - 1} \right)\left( {1 - \frac{2M}{N}} \right)^{2t} + 1 \\ 
 \,\,\,\,\,\,\,\,\,\,\,\,\,\,\,\,\,\, = \left( {\frac{M}{N} - 1} \right)\left( {1 - 
\frac{2M}{N}} \right)^{2t}\left( {\frac{4M}{N}\left( {\frac{M}{N} - 1} 
\right) + 1} \right) + 1 \\ 
 \,\,\,\,\,\,\,\,\,\,\,\,\,\,\,\,\,\, = \left( {\frac{M}{N} - 1} \right)\left( {1 - 
\frac{2M}{N}} \right)^{2\left( {t + 1} \right)} + 1 \\ 
 \end{array}
\end{equation}

\end{itemize}
\end{proof}

\end{document}